\documentclass[12pt]{iopart}
\usepackage{graphicx}
\usepackage{color}
\newcommand{\ltsimeq}{\raisebox{-0.6ex}{$\,\stackrel
        {\raisebox{-.2ex}{$\textstyle <$}}{\sim}\,$}}
\newcommand{\gtsimeq}{\raisebox{-0.6ex}{$\,\stackrel
        {\raisebox{-.2ex}{$\textstyle >$}}{\sim}\,$}}

\begin{document}
\title[Bright solitary waves and trapped solutions of attractive BECs]{Bright solitary waves and trapped solutions in Bose-Einstein condensates with attractive interactions.}
\author{N.G. Parker$^{1}$,  S. L. Cornish$^2$, C. S. Adams$^2$ and A. M. Martin$^1$}
\address{$^1$ School of Physics, University of Melbourne, Parkville,
Victoria 3010, Australia
\\$^2$ Department of Physics, Durham University, Durham, DH1 3LE, UK}

\begin{abstract}
We analyse the static solutions of attractive Bose-Einstein
condensates under transverse confinement, both with and without
axial confinement. By full numerical solution of the
Gross-Pitaevskii equation and variational methods we map out the
condensate solutions, their energetic properties, and their critical
points for instability. With no axial confinement a bright solitary
wave solution will tend to decay by dispersion unless the
interaction energy is close to the critical value for collapse. In
contrast, with axial confinement the only decay mechanism is
collapse. The stability of a bright solitary wave solution increases
with higher radial confinement. Finally we consider the stability of
dynamical states containing up to four solitons and find good
agreement with recent experiments.
\end{abstract}

\maketitle

The presence of attractive interactions in Bose-Einstein condensates
(BECs) leads to rich and intriguing nonlinear phenomena. A key
example is the formation of bright soliton-like structures
\cite{strecker,khaykovich,cornish_new}. A bright soliton is a
one-dimensional (1D) density wave that propagates without spreading
due to a balance between attractive interactions and dispersion. In
3D and under transverse confinement, the analog is a bright solitary
wave (BSW) solution, which is self-trapped in the axial direction
\cite{perez_gaussian,perez_garcia,carr2,salasnich}.  Due to their
self-trapped nature, BSWs hold significant advantages for
atom-optical applications, such as atom interferometry. Another
important property of attractive BECs in 3D is the collapse
instability. In 3D a homogeneous condensate with attractive
interactions is always unstable to collapse \cite{nozieres}.
However, the presence of trapping can stabilise the condensate
against collapse up to a critical number of atoms
\cite{bradley,roberts}. Indeed, the collapse instability has been
crucial in the experimental formation of BSWs
\cite{strecker,khaykovich,cornish_new}. A highly-populated
repulsively-interacting BEC has its interactions switched to
attractive via a Feshbach resonance.  This induces the collapse of
the condensate, with one \cite{khaykovich} or more
\cite{strecker,cornish_new} BSWs emerging from the collapse. For the
case of multiple BSWs, a $\pi$-phase difference leads to a repulsive
solitonic interaction that is important in stabilising their
collisions against collapse \cite{salasnich,carr,carr3,phase}.

\begin{table}[b]
\caption{Parameters of the three matter wave `solitons' experiments
to date. For the ENS experiment, we use an estimated atom number of
$N=4500$, which falls within the error of their measurements, while
for the JILA experiment we assume N=1500. The trap ratio $\lambda$
is defined as $\lambda=\omega_z/\omega_r$.}
\begin{tabular*}{\textwidth}{@{}l*{15}{@{\extracolsep{0pt plus12pt}}l}}
\br Experiment & Atomic species & $\omega_r/2\pi$ (Hz) & $|\lambda|$
& $a_{\rm s}$
(nm) & N & $k$\\ \mr Rice \cite{strecker} & $^7$Li & 800 & 0.005 (confining) & -0.16 & 5000 & 0.6 \\
ENS \cite{khaykovich} & $^7$Li & 710& 0.1 (expulsive) & -0.21 & 4500 & 0.65 \\
JILA \cite{cornish_new} & $^{85}$Rb & 17.5 & 0.4 (confining) &
-0.6 & 1500 & 0.35\\
\br
\end{tabular*}
\end{table}

Table 1 summarises the three experiments to date that have generated
BSWs of attractive BECs, at Rice University \cite{strecker}, ENS in
Paris \cite{khaykovich}, and JILA \cite{cornish_new}.  They feature
cylindrically-symmetric traps with radial harmonic confinement of
frequency $\omega_r$, and either a confining or expulsive axial
harmonic potential. Note that due to the presence of axial
confinement these are not true solitonic states and from now on we
will generally define BSWs to be solutions under zero axial
confinement. The strength of the atomic interactions, characterised
by the {\it s}-wave scattering length $a_{\rm s}$ ($a_{\rm s}<0$ for
attractive interactions), relative to the trapping potential is
crucial in determining the onset of collapse, and can be
parameterised in terms of the dimensionless parameter \cite{k_def},
\begin{equation}
k=\frac{N|a_{\rm s}|}{a_r}, \label{eqn:k}
\end{equation}
where $a_r=\sqrt{\hbar/m\omega_r}$ is the radial harmonic oscillator
length and $m$ is the atomic mass. Collapse occurs when the atom
number $N$ exceeds a critical population $N_{\rm c}$ or,
correspondingly, when the interaction parameter $k$ exceeds a
critical value $k_{\rm c}$.  The value of $k_{\rm c}$ is a crucial
consideration and has been the subject of several theoretical
investigations, both for condensates under a 3D potential
\cite{perez_gaussian,carr,ruprecht,
k_sph_sym,gammal,gammal2,yukalov} and for true BSWs under zero axial
confinement \cite{carr2,salasnich}. It depends on the geometry of
the trap and is of the order of unity. An accurate method to probe
$k_{\rm c}$ is to numerically solve the full 3D Gross-Pitaevskii
equation that describes the BEC mean-field, and isolate the point
where solutions no longer exist
\cite{carr2,ruprecht,gammal,gammal2}. This method is
numerically-intensive but can be approximated using a variational
approach \cite{perez_gaussian,carr2}. Indeed, the variational method
has been used to analyse solutions over the full range of axial trap
geometries - confining, expulsive and zero axial trapping.  However,
this has not been directly compared with the more accurate method of
solving the full Gross-Pitaevskii equation. In addition to these
methods, the nonpolynomial Gross-Pitaevskii equation, an effective
1D equation derived from the 3D Gross-Pitaevskii equation, has been
used to predict the critical interaction parameter for a bright
solitary wave \cite{salasnich}, while a simple analytic approach
\cite{yukalov} has been used to derive the critical interaction
parameter for an attractive condensate under 3D trapping.

In this paper we theoretically analyse the solutions of an
attractive BEC featuring radial confinement and under the full range
of axial potentials - no axial trapping, confining axial potential
and expulsive axial potential.  This is performed by solving the
full Gross-Pitaevskii equation and employing an approximate
variational approach. We map the regimes of instability and the
energetics of the solutions. In Section 1 we present our theoretical
framework of the Gross-Pitaevskii equation and the variational
approach. In Section 2 we use these methods to examine the stable
BSW solutions in the absence of axial trapping and the occurrence of
a finite excitation energy to the solution. In Section 3 we consider
the solutions under an axial potential, which is either confining or
expulsive, and compare to experiment. Finally in Section 4 we model
the multiple interacting `solitons' observed in the JILA experiment
\cite{cornish_new} and probe the critical number for such states,
finding good agreement between experiment and theory.

\section{Theoretical framework}
\subsection{Gross-Pitaevskii equation}

In the limit of zero temperature a dilute BEC can be approximated by
a mean-field `wavefunction' $\psi({\bf r},t)$ which satisfies the
Gross-Pitaevskii equation \cite{pethick},
\begin{equation}
i\hbar \frac{\partial \psi}{\partial
t}=\left[-\frac{\hbar^2}{2m}\nabla^2 + \frac{1}{2}m\omega_r ^2\left(
r^2 + \lambda^2 z^2\right)+gN|\psi|^2 \right]\psi.
\end{equation}
Here $N$ is the number of atoms in the condensate and $g=4\pi
\hbar^2 a_{\rm s}/m$ parameterises the strength of the {\it s}-wave
atomic interactions. The trap used to confine the BEC is assumed to
be harmonic and cylindrically-symmetric, where $\omega_r$ is the
transverse trap frequency and $\lambda=\omega_z/\omega_r$ is the
trap ratio. The wavefunction has been normalised such that $\int
|\psi({\bf r})|^2 d^3{\bf r}=1$. Furthermore the energy of the
system is defined by the Gross-Pitaevskii energy functional,
\begin{eqnarray}
E[\psi]&=&\int d^3 {\bf r} \left\{\frac{\hbar^2}{2m}|\nabla
\psi({\bf r})|^2+\frac{1}{2}m\omega_r ^2 \left( r^2 + \lambda^2
z^2\right)|\psi({\bf r})|^2+\frac{gN}{2}|\psi({\bf r})|^4 \right\}.
\label{eqn:E_func}
\end{eqnarray}
The basic Gross-Pitaevskii equation provides an excellent model of
BECs at the mean-field level \cite{minguzzi}, and although it is
insufficient to describe collapse dynamics, where higher order
effects such as three-body loss become considerable, it provides a
good model to infer the {\em onset} of collapse
\cite{ruprecht,gammal}.  For finite $g$ there are no analytic
solutions of the 3D Gross-Pitaevskii equation and we obtain the
solutions numerically via the imaginary-time technique: by
propagating the Gross-Pitaevskii equation in imaginary time
($t\rightarrow-it$), the system relaxes to the lowest energy state
of the system \cite{minguzzi}.

\subsection{Variational method}
\label{sec:ansatz}

As well as solving the Gross-Pitaevskii equation numerically we use
a variational technique to derive approximate solutions and give
valuable insight without intensive numerics. Assuming tight radial
trapping we decouple the 3D wavefunction $\psi({\bf r})$ into the
product of an axial and radial component, with the radial component
assumed to be a gaussian harmonic oscillator state
\cite{perez_gaussian,perez_garcia,carr2,salasnich}. For the axial
component there are two regimes:

(i) For $|\lambda| \ll 1$ the axial direction is dominated by
interactions. In 1D and the limiting case of $\lambda=0$, the
Gross-Pitaevskii equation supports bright soliton solutions of the
form $\psi(z)=\sqrt{2\xi}{\rm sech}(z/\xi)$ \cite{pethick}. The
healing length $\xi=\hbar/\sqrt{n_{\rm 1D}m|g|}$ characterises the
soliton width, where $n_{\rm 1D}$ is the peak 1D density.  Assuming
our 3D solution to have this axial profile leads to the ansatz
\cite{perez_garcia,carr2,salasnich},
\begin{equation}
\psi_{\rm sech}(r,z)=\left(\frac{1}{2\pi l_z l_r^2}\right)^{1/2}{\rm
sech}\left( \frac{z}{l_z}\right)\exp \left( -\frac{r^2}{2l_r^2}
\right). \label{eqn:BSWansatz}
\end{equation}
where $l_z$ and $l_r$ represent the axial and radial sizes,
respectively. Substituting this ansatz into the energy functional of
equation~(\ref{eqn:E_func}) leads to the ansatz energy \cite{carr2},
\begin{equation}
E_{\rm
sech}=\frac{\hbar^2}{2m}\left(\frac{1}{l_r^2}+\frac{1}{3l_z^2}\right)+
\frac{1}{2}m\omega_r^2\left(l_r^2+\frac{\pi^2\lambda^2l_z^2}{12}\right)+\frac{a_{\rm
s}N\hbar^2}{3ml_z l_r^2}. \label{eqn:BSWland}
\end{equation}

(ii) For $|\lambda| > 1$, the harmonic potential will dominate the
axial direction.  For a confining axial potential ($\lambda^2>0$),
it is then more appropriate to assume a gaussian axial profile.
This leads to,
\begin{equation}
\psi_{\rm gaus}(r,z)=\left(\frac{1}{\pi^{3/2} l_z
l_r^2}\right)^{1/2}\exp \left( -\frac{z^2}{2l_z^2} \right)\exp
\left( -\frac{r^2}{2l_r^2} \right), \label{eqn:HOansatz}
\end{equation}
and,
\begin{equation}
E_{\rm
gaus}=\frac{\hbar^2}{2m}\left(\frac{1}{l_r^2}+\frac{1}{3l_z^2}\right)+
\frac{1}{2}m\omega_r^2\left(l_r^2+\frac{\lambda^2l_z^2}{2}\right)+\frac{a_{\rm
s}N\hbar^2}{\sqrt{2\pi}m l_z l_r^2}. \label{eqn:HOland}
\end{equation}

Equations~(\ref{eqn:BSWland}) and (\ref{eqn:HOland}) define an
energy landscape in terms of $l_z$ and $l_r$.  Minimising the energy
with respect to these variational parameters leads to the
variational solution.

\section{Bright solitary wave solutions in the absence of axial
trapping} \label{sec:axial_homo}
\subsection{Stable and unstable energy landscapes}

\begin{figure}[t]
\begin{center}
\includegraphics[width=14cm]{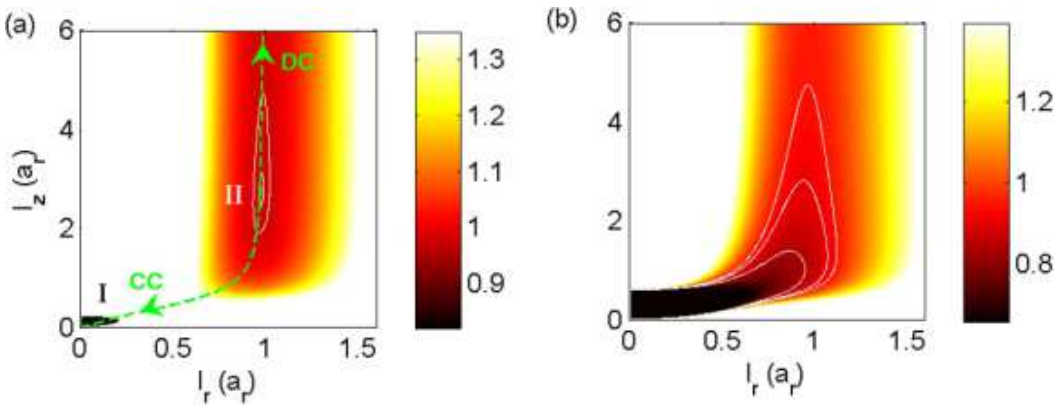}
\end{center}
\caption{Energy landscape according to equation~(\ref{eqn:BSWland})
for $\lambda=0$ in the (a) stable regime $k<k_{\rm c}$ $(k=0.35)$
and (b) unstable regime $k>k_{\rm c}$ ($k=0.9$). Energy is in units
of $\hbar\omega_r$. White contours highlight the shape of the
landscapes. In (a) two key regions are indicated: (I) the global
energy minimum (collapse region) and (II) the local energy minimum
corresponding to the variational solution. Whereas (a) features two
minima, corresponding to collapse and the BSW solution, (b) features
only a single minima representing collapse. In (a) the dashed green
line indicates the low-energy path through the energy landscape from
the origin to large $l_z$. For $l_z<l_z^0$ this path represents the
collapse channel (CC) and for $l_z>l_z^0$ it represents the
dispersive channel (DC).} \label{fig1}
\end{figure}

\begin{figure}[t]
\begin{center}
\includegraphics[width=8cm,clip=true]{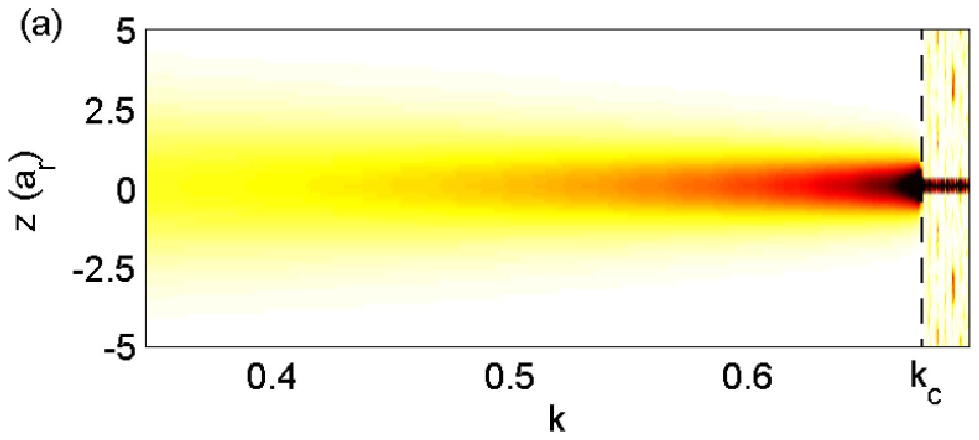}
\hspace{0.3cm}
\includegraphics[width=5.5cm,clip=true]{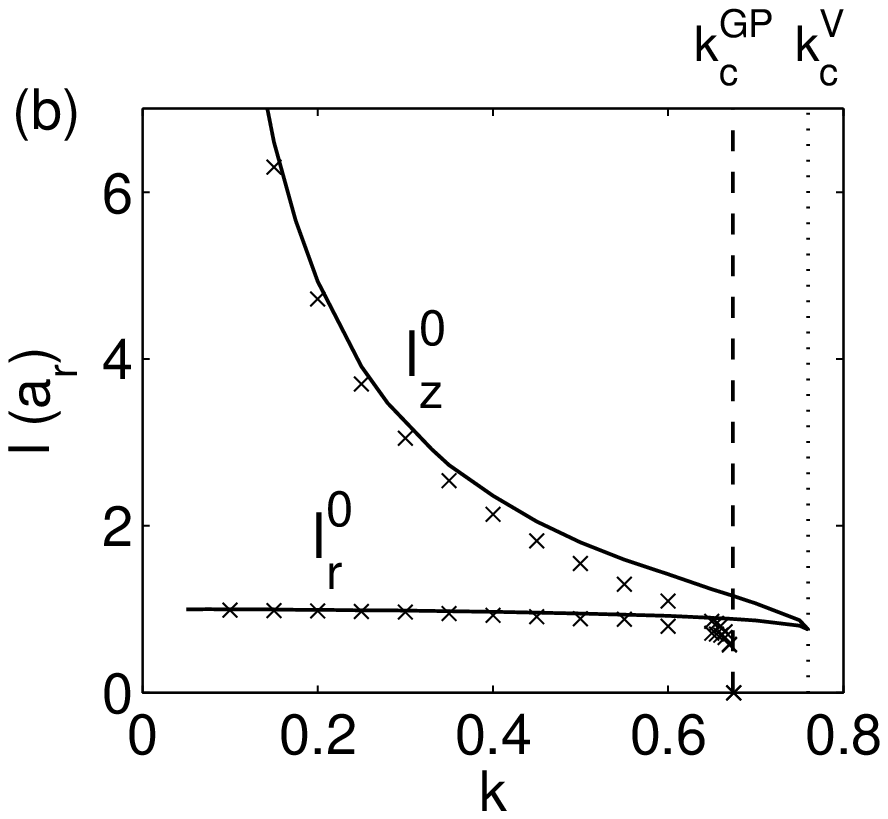}
\end{center}
\caption{(a) Evolution of the axial density (integrated over the
radial direction) of a BSW as the interaction parameter $k$ is
adiabatically increased past the critical point for collapse $k_{\rm
c}$.  The ramping rate is $dk/dt=0.12~{\rm s}^{-1}$. (b) Axial and
radial width of the BSW as a function of interaction strength $k$
according to the 3D Gross-Pitaevskii equation (crosses) as it
evolves under adiabatic ramping. The predictions from the sech-based
variational method of equations (\ref{eqn:BSWansatz}) and
(\ref{eqn:BSWland}) are shown by solid lines. The critical
interaction strength for collapse according to the sech-based
variational method $k^{\rm V}_{\rm c}$ and the Gross-Pitaevskii
equation $k^{\rm GP}_{\rm c}$ are indicated by vertical dotted and
dashed lines, respectively. } \label{fig2}
\end{figure}

We first examine BSW solutions (in the absence of axial trapping)
using the variational ansatz of equation (\ref{eqn:BSWansatz}). This
approach been performed previously by Carr and Castin \cite{carr2},
while Perez-Garcia {\it et al.} \cite{perez_gaussian} performed a
similar analysis using the gaussian ansatz of
equation~(\ref{eqn:HOansatz}). A typical energy landscape according
to equation (\ref{eqn:BSWland}) for a stable BSW solution is shown
in figure~\ref{fig1}(a), corresponding to $k=0.35$. At the origin
(region I) the interaction term in equation~(\ref{eqn:BSWland})
diverges to negative values since $a_{\rm s}<0$ for attractive
condensates. This region represents the physical collapse of a
wavepacket, although the global energy minimum itself is unphysical
since it has zero width \footnote{In reality the collapse stops at a
finite width when the three-body losses become considerable and
reduce the atom number to below the critical number
\cite{minguzzi}.}. Importantly, there exists a stable local energy
minimum indicated by region II in figure~\ref{fig1}(a). This
represents the BSW solution. A typical unstable energy landscape is
shown in figure~\ref{fig1}(b), for a large interaction parameter of
$k=0.9$. No local energy minimum exists, and the whole parameter
space is unstable to collapse.

\subsection{Critical point for collapse}
\label{sec:crit_BSW}

Using the sech-based ansatz of equation (\ref{eqn:BSWansatz}), Carr
and Castin have shown that the critical interaction strength for
collapse of a BSW is $k^{\rm V}_{\rm c}=0.76$ \cite{carr2}.
Similarly, Perez-Garcia {\it et al.} \cite{perez_gaussian} have used
the gaussian-based ansatz of equation (\ref{eqn:HOansatz}) to
predict a critical point of $k^{\rm V}_{\rm c}=0.778$.  Using the
non-polynomial Gross-Pitaevskii equation, Salasnich {\it et al.}
predict that $k^{\rm V}_{\rm c}=2/3$ \cite{salasnich}.  These
methods consider approximated solutions to the 3D Gross-Pitaevskii
equation. A more exact method, although considerably more intensive,
is to numerically solve the full 3D Gross-Pitaevskii equation
\cite{carr2,gammal}.

In order to isolate $k_{\rm c}$ from the full 3D Gross-Pitaevskii
equation we employ an adiabatic ramping technique. Firstly, a stable
BSW solution ($k \ll k_{\rm c}$) is obtained by propagating the 3D
Gross-Pitaevskii equation in imaginary time. Then, in real time, the
interaction parameter $k$ is increased adiabatically. We employ a
ramping rate $dk/dt=0.12~ {\rm s}^{-1}$. We have verified that a
lower ramping rate does not change the point of collapse. An example
of a BSW under adiabatic ramping is shown in figure \ref{fig2}(a).
For $k<k_{\rm c}$ the condensate progresses through the stationary
state solutions, becoming progressively narrower and of higher
density, but at $k=k_{\rm c}$ the condensate suddenly collapses.
Using this technique we have isolated the critical interaction
strength according to the full Gross-Pitaevskii equation for a BSW
to be $k^{\rm GP}_{\rm c}=(0.675\pm0.005)$. The error margin arises
from the finite time over which the wavepacket collapses. We have
confirmed that if the value of $k$ is ramped up to, and then
maintained at, any value less than this $k^{\rm GP}_{\rm c}$, the
condensate remains stable. This value is different to the value of
$k^{\rm GP}_{\rm c}=0.627$ obtained by Carr and Castin \cite{carr2}
through numerical relaxation of the 3D Gross-Pitaevskii equation.
However, our result is consistent with Gammal {\it et al.}
\cite{gammal} who employed numerical relaxation under a weak axial
trap of $\lambda=0.01$ (very close to the axially-homogeneous case)
and obtained $k^{\rm GP}_{\rm c}=0.676$.  Furthermore, this value is
in good agreement with Salasnich {\it et al.} \cite{salasnich} who
analytically derived $k_{\rm c}=2/3$ from the nonpolynomial
Gross-Pitaevskii equation.

In contrast to the adiabatic ramping method of quantifying $k^{\rm
GP}_{\rm c}$, the numerical relaxation approach relies on searching
for solutions to the Gross-Pitaevskii equation (by propagating in
imaginary time) for increasing values of $k$, and locating $k^{\rm
GP}_{\rm c}$ as being the point when solutions can no longer be
found. This has been employed successfully to obtain the critical
interaction strength for attractive condensates under 3D trapping
\cite{ruprecht,gammal}. However, for BSWs close to the critical
interaction strength, the local energy minimum representing the BSW
solution can be so shallow (see solid line in figure~\ref{fig3}(c))
and localised that the imaginary time method is very sensitive to
the initial guess and can become ineffective unless the initial
guess is practically the solution itself.

\subsection{Lengthscales of the bright solitary wave}

The axial and radial widths of the BSW solutions, $l^0_z$ and
$l^0_r$, are shown in figure~\ref{fig2}(b) as a function of $k$,
from the full Gross-Pitaevskii equation (crosses) and the sech-based
variational method (lines).  We evaluate the axial width of the
Gross-Pitaevskii equation solutions as being the lengthscale over
which the axial density decreases to ${\rm sech}^2(1)n_0\simeq 0.42
n_0$, where $n_0$ is the central peak density, and the radial width
as the lengthscale over which the radial density decreases to
$n_0/e$.

The variational and full Gross-Pitaevskii equation methods give the
same qualitative description. The wave is elongated in the axial
direction ($l^0_z>l^0_r$), with radial width $l^0_r$ remaining close
to $a_r$ while the axial width $l^0_z$ diverges as $k\rightarrow0$.
For increasing interaction strength, $l_z$ approaches $l_r$, and at
the point of collapse the BSW is approximately isotropic
($l^0_r=l^0_z$), as noted elsewhere \cite{carr2}. However, the
variational method consistently overestimates the widths predicted
by the Gross-Pitaevskii equation. This implies that the variational
solutions have lower peak density than the Gross-Pitaevskii equation
solutions and explains why they predict a higher value of $k_{\rm
c}$.

\subsection{BSW energetics}
\begin{figure}[t]
\begin{center}
\includegraphics[width=11cm,clip=true]{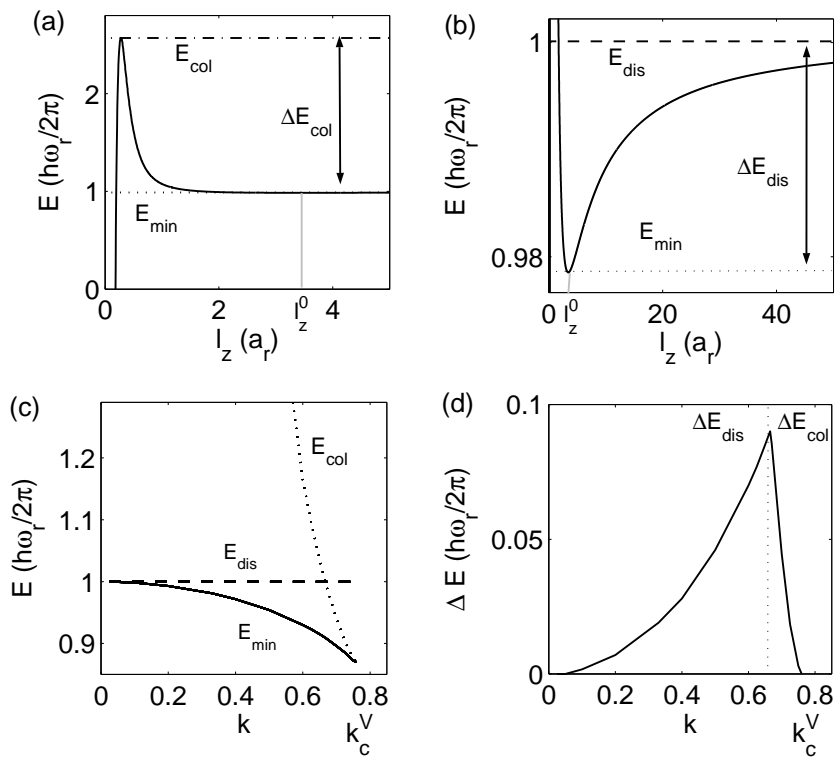}
\end{center}
\caption{Energetics of the BSW (for $\lambda=0$) according to the
sech-based ansatz. (a) Energy profile along the low-energy path
through the energy landscape (dashed line in figure \ref{fig1}(a)).
The interaction strength is $k=0.35$. The energy of the solution
$E_{\rm min}$ (dotted line), the energy barrier to the collapse
region $E_{\rm col}$ (dot-dashed line) and the resulting energy
difference $\Delta E_{\rm col}$ are indicated. The position of the
energy minimum $l_z^0$ is shown (grey line). (b) Same as (a) but
plotted on a different scale to highlight the energy of the
dispersive channel $E_{\rm dis}$ (dashed line) and the resulting
energy difference $\Delta E_{\rm dis}=E_{\rm dis}-E_{\rm min}$. (c)
The variation of $E_{\rm min}$ (solid line), $E_{\rm col}$ (dotted
line) and $E_{\rm dis}$ (dashed line) with the interaction parameter
$k$. (d) The excitation energy $\Delta E$ of the solution, i.e. the
lower of either the collapse excitation energy $\Delta E_{\rm col}$
or dispersive excitation energy $\Delta E_{\rm dis}$.} \label{fig3}
\end{figure}

Consider the variational BSW solution indicated by region II in
figure \ref{fig1}(a).  We denote the energy minimum in this region
by $E_{\rm min}$. In the radial direction, the local energy minimum
is well-bounded due to the dominance of the radial confinement.  In
the axial direction, however, there is low energy path in the energy
landscape which leads from the origin to large $l_z$, passing
through the local BSW energy minimum. We have indicated this path on
the energy landscape in figure \ref{fig1}(a) by the dashed green
line. This low-energy path represents the most
energetically-favourable states of the BSW at a given axial width.
In figure \ref{fig3}(a) and (b) we show the variation of the energy
along this path, plotted on different scales to highlight
contrasting features. For $l_z<l_z^0$, the energy (figure
\ref{fig3}(a)) initially increases with decreasing $l_z$ due to the
growth of the kinetic terms in equation (\ref{eqn:BSWland}).
However, for even smaller $l_z$ the negative interaction energy
begins to dominate and causes the energy to decrease (and ultimately
diverge to $-\infty$ as $l_z\rightarrow 0$). Importantly, this gives
rise to a finite energy barrier which stabilises the system against
collapse. For $l_z>l_z^0$, the energy (figure \ref{fig3}(b))
increases weakly with $l_z$.  This is because interaction energy in
equation (\ref{eqn:BSWland}) falls off more slowly than the kinetic
energy. This energy profile means that if one creates a state with
$l_z>l_z^0$, then the interactions are sufficient to overcome
dispersion and the state will oscillate in the energy minimum, with
its length oscillating around $l^0_z$.

In the limit $l_z \rightarrow \infty$, the axial kinetic and
interaction energy terms in equation (\ref{eqn:BSWland}) tend
towards zero and the energy landscape becomes $E_{\rm sech}(l_r,l_z
\rightarrow \infty)=\hbar^2/(2ml_r^2)+m\omega_r^2 l_r^2/2$, with a
minimum energy $E_{\rm dis}=\hbar \omega_r$ at $l_r=a_r$. As a
result we see that the low energy channel in figure \ref{fig3}(b)
tends asymptotically towards $E_{\rm dis}$ as $l_z\rightarrow
\infty$. The fact that the energy is finite at $l_z=\infty$ has
important consequences. If one adds {\em additional} axial kinetic
energy to the system, e.g. by exciting an axial breathing mode, we
can form states with higher energy than described by equation
(\ref{eqn:BSWland}), i.e. states that lie above the line in figure
\ref{fig3}(a) and (b). Importantly, if the energy of this excited
state is greater than $E_{\rm dis}$ then it has sufficient energy to
overcome the interactions and expand indefinitely.  This means that
there is a finite excitation energy $\Delta E_{\rm dis}=E_{\rm
dis}-E_{\rm min}$ to induce dispersion of the BSW, as indicated in
figure \ref{fig3}(b).  We term this the {\it dispersive channel}
(DC), as illustrated in figure \ref{fig1}(a).

There is another channel by which the BSW can be excited to a
non-solitonic state.  For axial widths less than $l_z^0$, there is a
high-energy saddle point in the energy surface that links the local
BSW energy minimum to the collapse region. We term this the {\it
collapse channel} (CC) and have indicated it in figure
\ref{fig1}(a). In figure \ref{fig3}(a) we see this feature as an
energy barrier. The energy maximum has amplitude $E_{\rm col}$ and
so presents an energy barrier of magnitude $\Delta E_{\rm
col}=E_{\rm col}-E_{\rm min}$ against the collapse of the BSW.

In figure \ref{fig3}(c) we show the variation of these energies as a
function of the interaction parameter $k$. The energy minimum
$E_{\rm min}$ decreases slightly with increasing $k$ (solid line).
The energy of the dispersive channel is independent of $k$ (dashed
line). The collapse channel energy $E_{\rm col}$ (dotted line)
decreases rapidly with increasing $k$. When the critical interaction
strength $k_{\rm c}$ is reached, $\Delta E_{\rm col}=0$ (there is no
longer an energy barrier to prevent collapse), and for $k>k^{\rm
V}_{\rm c}$ no variational BSW solutions exist.

These channels introduce a finite excitation energy to excite the
BSW to a non-solitary state. The minimal excitation energy (for
either of these channels), $\Delta E$, characterises the stability
of the BSW. This excitation energy is plotted in
figure~\ref{fig3}(d) as a function of $k$. For $k \ltsimeq 0.66$,
the dispersive channel is of lowest energy, with $\Delta E$
increasing with $k$. If the BSW were excited in this regime it would
be prone to dispersion. However, for $k \gtsimeq 0.66$ the collapse
channel is of lowest energy and $\Delta E$ rapidly decreases to zero
as $k\rightarrow k^{\rm V}_{\rm c}$. A BSW excited in this regime
will be prone to collapse.
The maximum excitation energy $\Delta E\approx 0.09 \hbar \omega_r$
is finite and occurs at $k \approx 0.66$. Consequently, to maximize
the thermal and dynamical stability of a BSW requires tight radial
confinement and $k\approx 0.66$.

According to the JILA parameters (but with $\lambda=0$), the weak
radial trapping leads to a small excitation energy $\Delta E\approx
0.08$~nK, whereas the strong radial trapping of the Rice and ENS
parameters leads to a much larger value of $\Delta E \approx 3$ nK.
In the experiments, following the collapse the resulting solitonic
states are sufficiently dilute that it is unlikely that the system
reaches thermal equilibrium within the experimental time frame.
However, we can estimate the thermal energy of the system based on
the temperature prior to collapse.  In the JILA experiment there are
initially approximately 15000 atoms in the BEC and 500 thermal
atoms, corresponding to a temperature of $4.6$~nK. This suggests
that a BSW in the JILA system will be unstable to thermal effects,
whereas the Rice and ENS systems may just support
thermodynamically-stable BSWs. However, BSWs may have enhanced
thermodynamic stability due to self-cooling via the radiation of hot
atoms in the untrapped direction \cite{carr3}. Of course, the
experiments themselves featured an axial potential, unlike the
$\lambda=0$ limit considered here, and we will see in the next
section that this modifies the thermodynamic stability dramatically.

\section{Solutions under an inhomogeneous axial potential}
\label{sec:trapping}

\subsection{Confining axial potential}

\begin{figure}[t]
\begin{center}
\includegraphics[width=14cm,clip=true]{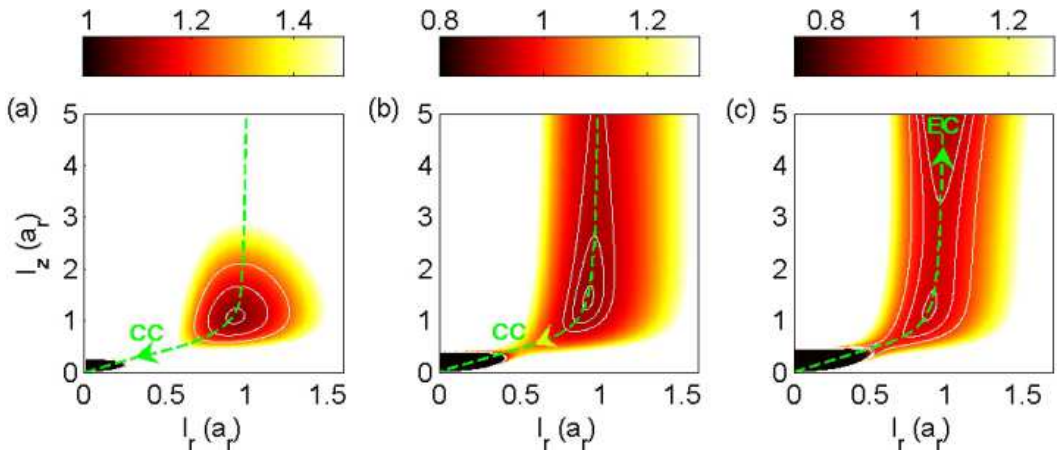}
\end{center}
\caption{Energy landscapes of equation~(\ref{eqn:BSWland}) for
attractive BECs under the confining and expulsive axial potentials
used experimentally: (a) JILA parameters with $k=0.35$ and confining
axial trapping $\lambda=0.4$; (b) Rice parameters with $k=0.6$ and
confining axial trapping $\lambda=0.005$.  (c) ENS geometry with
$k=0.65$ and an expulsive axial potential ($\lambda^2<0$) with trap
ratio $|\lambda|=0.1$. Energy is presented in units of
$\hbar\omega_r$ and white contours highlight the shape of the
landscapes.  The dashed line in each plot indicates the low energy
pathway through the energy landscape, from the origin to large
$l_z$. In (a) and (b) the preferred decay route is the collapse
channel (CC) while in (c) it is the expansive channel (EC).}
\label{fig4}
\end{figure}

\begin{figure}[t]
\begin{center}
\includegraphics[width=12cm,clip=true]{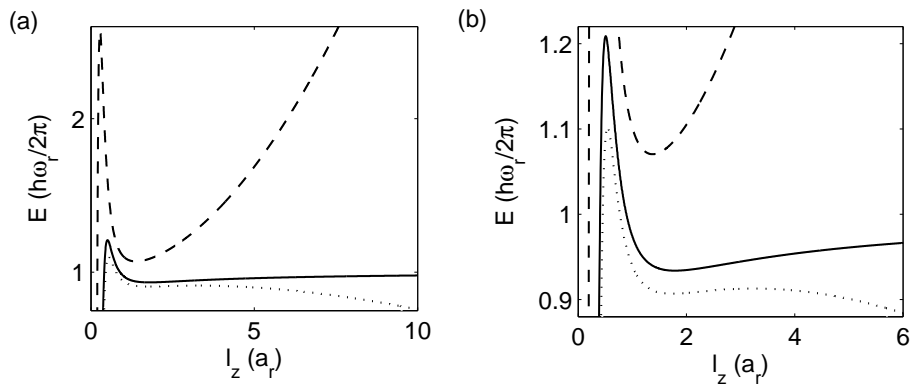}
\end{center}
\caption{(a) Energy profile along the low-energy path through the
energy landscape (dashed lines in figure \ref{fig4}(a)-(c)) for the
JILA system (dashed line), Rice system (solid line) and ENS system
(dotted line). (b) Same as (a) but on a different scale to highlight
the small-scale features for the Rice and ENS system.  Although not
evident on the lengthscales shown, the energy in the Rice system
tends to infinity as $l_z\rightarrow \infty$ due to the effect of
the confining potential. Energy is presented in units of
$\hbar\omega_r$ and each experiment features a different $\omega_r$.
Although the collapse barrier in the Rice system appears smaller
than for the JILA system, it is in reality much larger due to the
much larger value of $\omega_r$ employed in the Rice experiment.}
\label{fig4b}
\end{figure}
We now consider the condensate solutions in the presence of a
confining axial potential ($\lambda^2>0$).  In
figure~\ref{fig4}(a)-(b) we plot the energy landscapes according to
equation~(\ref{eqn:BSWland}) for the parameters employed in the JILA
and Rice experiments.  In each case, the lowest energy path through
the landscape is illustrated by dashed lines.  In figure \ref{fig4b}
we plot the energy profile along the low-energy path through the
energy landscape. Due to the axial confinement the $\lambda^2
l_z^2$-term in equation~(\ref{eqn:BSWland}) dominates for large
$l_z$ and ultimately leads to the energy increasing with $l_z$.
Consequently there is no dispersive channel in this system, as
evident from figure \ref{fig4b} (dashed line). The presence of even
small axial trapping therefore has a profound increase in the
stability of the state. The large trap ratio of the JILA experiment
($\lambda=0.4$) significantly modifies the energy landscape from the
$\lambda=0$ case (figure~\ref{fig1}(a)), whereas the weak trap ratio
of the Rice experiment ($\lambda=0.005$) only has a weak effect on
the energy landscape (on the lengthscales shown).

It is still possible to excite the solution out of the energy basin
into the collapse region, as indicated by arrows in figure
\ref{fig4}(a) and (b). According to the sech-based variational
method, this excitation energy is $\Delta E=1.32\hbar
\omega_r\approx 1$ nK for the JILA experimental parameters and
$\Delta E=0.24\hbar \omega_r\approx 9$ nK for the Rice experimental
parameters, which are of the order of the typical thermal energy of
the BEC. These values are significantly larger than the $\lambda=0$
case and indicate enhanced stability under an axial trap.

\begin{figure}[t]
\begin{center}
\includegraphics[width=11 cm,clip=true]{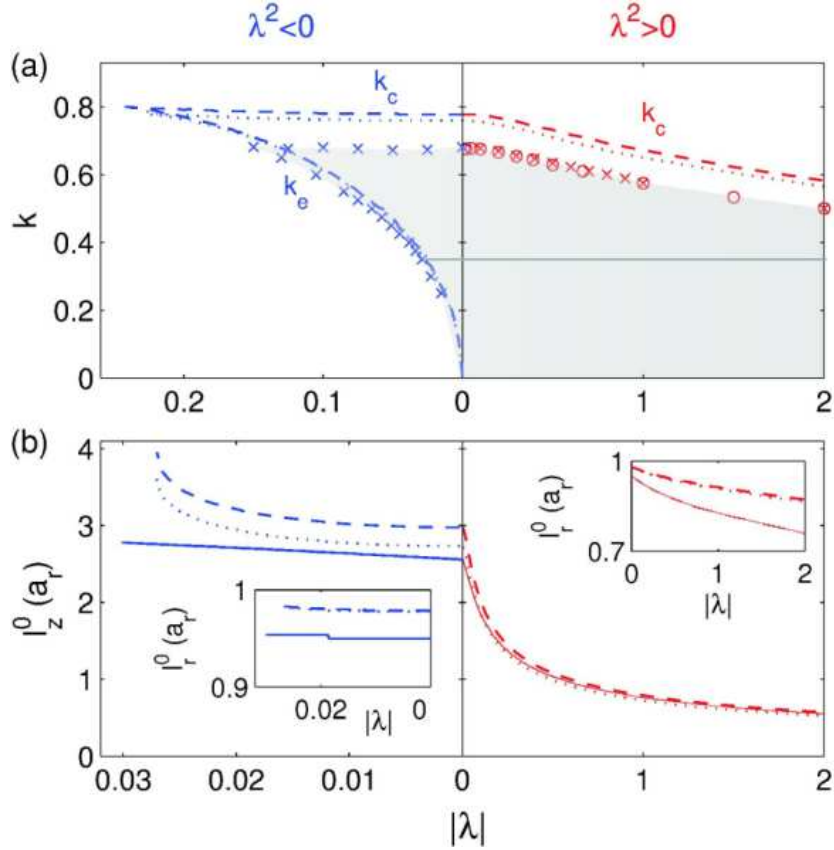}
\end{center}
\caption{(a) Critical interaction strengths as a function of
$|\lambda|$ for a confining potential $\lambda^2>0$ (red data,
right-hand side of figure) and expulsive potential $\lambda^2<0$
(blue data,left-hand side of figure). The region of stable
solutions, according to the Gross-Pitaevskii equation, is shaded.
For the confining potential there is only single critical point,
namely $k_{\rm c}$ for collapse, while for the expulsive trap there
is additional critical point for expansion $k_{\rm e}$.  Our
Gross-Pitaevskii equation results are shown by crosses, the
sech-based variational method by the dotted line and the
gaussian-based variational method by the dashed line. The
predictions of Gammal {\it et al.} \cite{gammal} for a confining
axial trap are shown (circles). (b) Axial size of the condensate
solutions as a function of $|\lambda|$ according to the
Gross-Pitaevskii equation (solid line), the sech-based variational
method (dotted line) and the gaussian-based variational method
(dashed line). We employ an interaction strength of $k=0.35$ such
that we probe the range of solutions shown by the horizontal solid
line in (a).  The insets show how the radial width depends on
$|\lambda|$. Note that the scale for $|\lambda|$ on the left and
right sides of (a) and (b) are different.} \label{fig5}
\end{figure}

For an attractive BEC in a spherically-symmetric trap the critical
interaction strength has been predicted to be $k_{\rm c}\approx
0.57$ by a variety of methods \cite{k_sph_sym,gammal,gammal2}. Using
imaginary-time relaxation of the Gross-Pitaevskii equation, Gammal
{\it et al.} \cite{gammal} predicted $k^{\rm GP}_{\rm c}$ for
various trap ratios and, for the JILA trap geometry, give excellent
agreement with the experimentally measured value of $k_{\rm c}$
\cite{roberts,claussen}.  By simulating the Gross-Pitaevskii
equation under the adiabatic ramping technique we have obtained the
critical interaction strength for collapse $k_{\rm c}$, with the
results presented in figure~\ref{fig5}(a) (crosses).  Note that the
shaded region represents stable solutions. For a confining axial
potential our Gross-Pitaevskii equation results are consistent with
the results of Gammal {\it et al.} \cite{gammal} (circles).  We also
present the results of the sech-based variational method (dotted
line) and the gaussian-based variational method (dashed line).
Although these predictions give the same qualitative features,
namely that $k_{\rm c}$ decreases weakly with increasing trap ratio
$\lambda$, these variational methods consistently overestimate the
critical interaction strength by around $20\%$.

In figure~\ref{fig5}(b) we plot the axial width $l^0_z$ of the
solutions as a function of the trap ratio $\lambda$ for a fixed
interaction strength of $k=0.35$. The Gross-Pitaevskii equation
(solid lines) and variational predictions (dotted and dashed lines)
show good quantitative agreement.  The radial widths $l^0_r$, shown
in the inset, also show reasonable agreement.

\subsection{Expulsive axial potential}

The ENS experiment employed an expulsive ($\lambda^2<0$) trap in the
axial direction \cite{khaykovich}.  This geometry leads to an
additional effect whereby the wavepacket can be ripped apart by the
expulsive trap \cite{khaykovich,carr2}. This occurs when $k$ is less
than a {\em lower} critical interaction strength for expansion
$k_{\rm e}$.  The upper and lower critical interaction strengths for
an expulsive trap have been mapped out using the sech-based
variational method \cite{khaykovich,carr2}.

The energy landscape for the ENS system is shown in
figure~\ref{fig4}(c).  The energy profile along the low-energy path
through the energy landscape is shown in figure \ref{fig4b} (dotted
line). The expulsive axial potential leads to a downward slope in
the energy landscape for large $l_z$. In this region wavepackets
will expand indefinitely. We term this the {\it expansive channel},
as indicated in figure \ref{fig4}(c).  Note that we define the
effect of expansion to be induced by the axial potential and so it
is distinct from the effect of dispersion discussed in Section
\ref{sec:axial_homo}. The energy barrier between the solution and
the expansive region {\it decreases} with decreasing interaction
strength, such that below $k_{\rm e}$ no solutions are stable
against expansion. Note that, according to the energy landscape
(figure~\ref{fig4}(c)), the minimum excitation energy is $\Delta
E\approx 0.006\hbar \omega_r\approx 0.2$ nK and corresponds to
excitation into the expansive region.  Although this thermodynamic
energy is less than the typical energy of a BEC, the expulsive trap
is expected to lead to the removal of hot atoms and so may lead to
anomalously low condensate temperatures.

We have calculated the critical interaction strengths for collapse
$k_{\rm c}$ and expansion $k_{\rm e}$ using the Gross-Pitaevskii
equation. In the former case we adiabatically ramp up $k$ at fixed
$|\lambda|$ until collapse in induced, and in the latter case we
adiabatically ramp up $|\lambda|$ at fixed $k$ until expansion is
observed. In figure~\ref{fig5}(a) (left-hand side) we plot $k_{\rm
c}$ and $k_{\rm e}$ as a function of trap ratio $|\lambda|$ using
the full Gross-Pitaevskii equation (crosses), and the sech-based
(dotted line) and gaussian-based (dashed line) variational methods.
Note that the range of trap ratios we present is an order of
magnitude smaller than for the case of the confining axial potential
(right-hand side of figure~\ref{fig5}(a)). The qualitative features
are that $k_{\rm c}$ (upper data) increases weakly with trap ratio
$|\lambda|$, while $k_{\rm e}$ (lower data) increases sharply from
zero at $|\lambda|=0$. The region between $k_{\rm c}(|\lambda|)$ and
$k_{\rm e}(|\lambda|)$ (shaded region in figure~\ref{fig5}(a))
represents stable solutions, the region above $k_{\rm c}(|\lambda|)$
represents interaction-induced collapse, and the region below
$k_{\rm e}(|\lambda|)$ represents potential-induced expansion. At
some critical trap ratio $|\lambda|_{\rm c}$, $k_{\rm c}=k_{\rm e}$,
and for $|\lambda|>|\lambda|_{\rm c}$ there are no stable solutions.
According to the variational methods, $|\lambda|^{\rm V}_{\rm
c}\approx 0.23$ while according to the Gross-Pitaevskii equation
$|\lambda|^{\rm GP}_{\rm c} \approx 0.15$.

\section{Dynamical multi-soliton states}
In the JILA experiment \cite{cornish_new} up to six long-lived
localised `solitons' \footnote{The experimental system is 3D and
features axial confinement, and so the wavepackets are not solitons
in the strict sense. However, in this section we will follow the
experimental protocol in referring to these wavepackets loosely as
solitons.} were observed. These wavepackets oscillated axially in a
robust manner, repeatedly colliding with each other. Interestingly,
the total number of atoms could greatly exceed the critical number
for collapse $N_{\rm c}$ and yet the dynamics remained stable.  This
is because each wavepacket contained less than $N_{\rm c}$ atoms and
so was individually stable to collapse, and furthermore a
$\pi$-phase difference existed between adjacent wavepackets to
prevent them coalescing and collapsing
\cite{carr2,salasnich,carr,phase,khawaja}. A $\pi$-phase difference
has also been inferred to exist in the train of solitons observed in
the Rice experiment \cite{strecker,salasnich,khawaja}. These
$\pi$-phase differences are thought to arise through the imprints of
quantum fluctuations \cite{khawaja} or the emergence of this stable
configuration through repeated instabilities \cite{carr,salasnich2}

The JILA experiment examined configurations of up to four distinct
soliton structures.  We will probe the stability of configurations
of several coexisting solitons in the specific JILA experimental
set-up, namely a $^{85}$Rb BEC in a three-dimensional trap geometry
$\omega_z=2\pi\times 6.8$ Hz and $\omega_r=2\pi\times 17.5$ Hz.
Crucially we will assume a $\pi$-phase difference between adjacent
solitons. A more rigorous approach to this problem would be to study
excited nonlinear states of the system rather than multiple
solitons, as has been performed in quasi-1D limit using Hermite
functions by Michinel {\it et al.} \cite{michinel}. Multi-soliton
states with alternating phase have also been shown to be supported
by the nonlinear Schr\"{o}dinger equation elsewhere
\cite{multisolitons}.


\subsection{Generation of dynamical multi-soliton states}

We impose $\pi$-phase differences to separate the attractive BEC
into several distinct wavepackets, and for each configuration
examine the critical number of atoms. We define the critical number
for $q$ distinct wavepackets to be $N_{\rm cq}$.  We generate the
$q=1$ solutions by imaginary time propagation of the
Gross-Pitaevskii equation with constant global phase and isolate the
critical number $N_{\rm c1}$ as when solutions can no longer be
found. In a similar manner, we form a $q=2$ state by solving in
imaginary time subject to a $\pi$-phase difference at $z=0$, and
infer the critical number $N_{\rm c2}$.

Generation of $q>2$ states is more complicated. To form a $q=3$
solution we employ two $\pi$-phase steps at $z=\pm L/2$, as
illustrated in figure~\ref{fig7}(a), to form three distinct regions
$z\leq-L/2$, $-L/2<z<+L/2$ and $z\geq+L/2$. Imaginary-time
propagation of the Gross-Pitaevskii equation under these phase
constraints can preferentially populate the central region to point
of collapse. Instead we generate the $q=3$ states dynamically, with
an example shown in figure~\ref{fig7}(c). Initially we form the
ground state with constant phase and repulsive interactions. Then,
in real time, the interactions are switched to attractive while the
stepped phase distribution (figure~\ref{fig7}(a)) is imprinted onto
the condensate for a finite time (a time of $\Delta t=80$ms works
effectively). This forms three wavepackets with no atom transfer
between them (figure~\ref{fig7}(c)). This dynamical method of
generating solitons is loosely analogous to the experimental
generation of solitons. However, whereas the alternating phases
arise naturally in the experiment, we must introduce them
numerically.

When the phase imprinting is terminated, and providing the critical
number $N_{\rm c3}$ is not exceeded, the solitons oscillate axially
and repeatedly collide (figure~\ref{fig7}(c)), as observed
experimentally.  The alternating phase structure prevents overlap
and suppresses the collapse instability.  However, the solitonic
interaction excite collective modes in the solitons.  When $N$
exceeds a critical value, the system is unstable to collapse and the
solitons are destroyed.  This critical value depends on the
lengthscale $L$, and so to obtain $N_{\rm c3}$ we must vary $L$ to
obtain the most stable configuration. We find that the most stable
configurations arise when $L\approx 2\xi$.  In order to generate a
$q=4$ state we perform a similar process by imposing three
$\pi$-phase steps at $z=0$ and $z=\pm L$, as illustrated in
figure~\ref{fig7}(b). Under stable parameters, four localised
solitons are formed, which subsequently oscillated in the system in
a stable manner.  An example is shown in figure~\ref{fig7}(d).

\subsection{Stability of dynamical multi-soliton states}
\begin{figure}
\begin{center}
\includegraphics[width=7cm,clip=true]{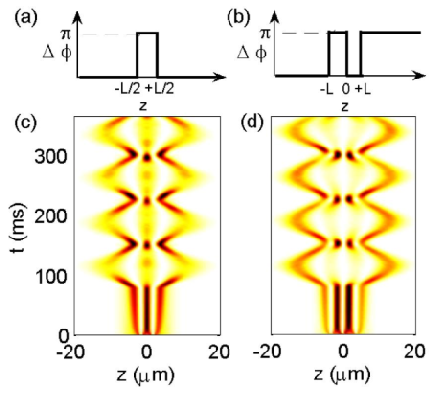}
\includegraphics[width=4.5cm,clip]{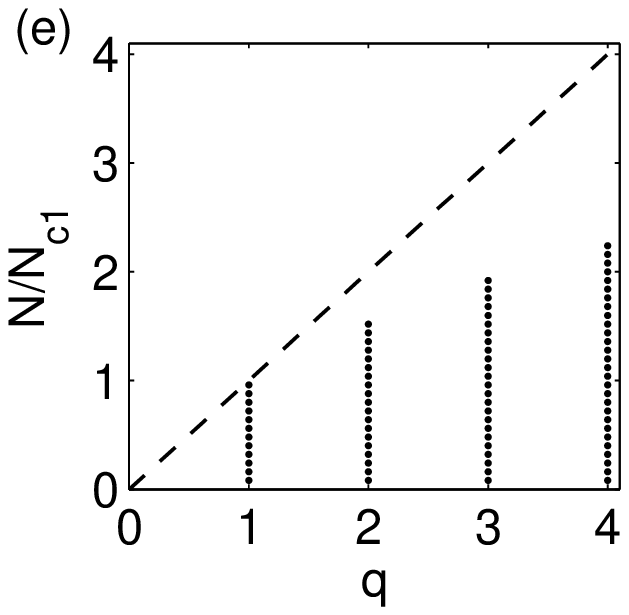}
\end{center}
\caption{(a)-(b) The stepped phase profiles, with characteristic
size $L$, used to dynamically generate $q=3$ and $q=4$ states. The
phase profiles are enforced until $t=80$ms. (c) Evolution of a $q=3$
and (d) $q=4$ state with $N=3000$, and $L=3.5\mu m$.  Following the
JILA experiment, we employ $a_{\rm s}=-0.6$ nm.  The plots show the
evolution of the axial density, integrated over the radial
direction.(e) Stability diagram of $N/N_{\rm c1}$ showing stable
configurations of up $q=1, 2, 3$ and $4$ states. The line $(N/N_{\rm
c})$ is plotted for comparison (dashed line).} \label{fig7}
\end{figure}

The stability of states with $q\leq4$ and a fixed scattering length
$a_{\rm s}=-0.6$nm is shown in figure~\ref{fig7}(e). If each soliton
was independent with critical number $N_{\rm c1}$,  the total
critical number for $q$ solitons would be $qN_{\rm c1}$. In
figure~\ref{fig7}(e) we have plotted this estimate as the ratio
$N/N_{\rm c}$ (dashed line).  However, the numerical data gradually
deviates from this line and $N_{\rm cq} \leq qN_{\rm c1}$,
indicating that the solitons are increasingly affected by each other
as their number increases.  A similar trend was predicted in the
quasi-1D approach of Michinel {\it et al.} \cite{michinel}.

\begin{figure}[t]
\begin{center}
\includegraphics[width=10cm]{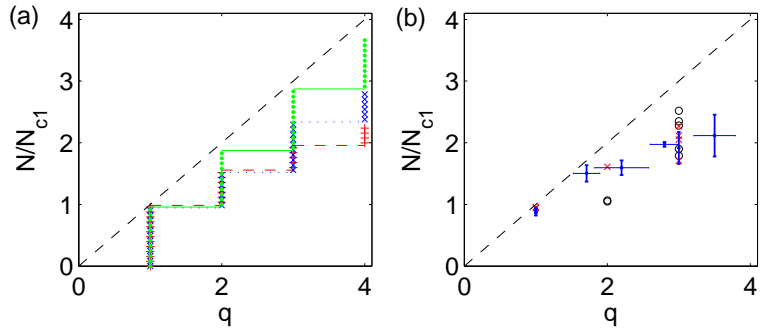}
\end{center}
\caption{(a) The ratio $N/N_{\rm c1}$  versus the most stable number
of solitons $q$ (up to $q=4$) for scattering lengths $a_{\rm
s}=-0.6$nm (red pluses/dashed line), $-0.77$nm (blue crosses/dotted
line), and $-1.59$nm (green circles/solid line). (b) The
experimental remnant data of \cite{cornish_new} (blue points with
error bars) and the corresponding numerical data where $N=N_{\rm
exp}$(black circles). Numerical data corresponding to $N=0.9N_{\rm
exp}$ is also shown (red crosses). Note that the points are for
different scattering lengths.} \label{fig8}
\end{figure}

An important question is how many solitons are favoured to exist for
a given number of atoms $N$. The addition of a soliton involves
adding a $\pi$-phase slip to the system, which is energetically
costly. We can therefore expect that the most
energetically-favourable number of solitons will be the minimum
number of solitons/phase slips that can support $N$ atoms without
being unstable to collapse. This data is plotted in
figure~\ref{fig8}(a) for the three scattering lengths employed in
the JILA experiment. All three data sets diverge away from the ratio
$N/N_{\rm c}$ as $q$ increases.  The divergence is greatest for low
scattering lengths, which is related to the fact that these
wavepackets are widest (since the soliton size is characterised by
$\xi\propto 1/\sqrt{a_{\rm s}}$) and therefore interact most with
their neighbours. The region above the points represents
configurations of the system which are completely unstable to
collapse. The region below the lines in figure~\ref{fig8}(a) also
represents stable configurations of the system, but these
configurations consist of a higher number of solitons/phase slips,
and therefore correspond to higher energy.

\subsection{Comparison with the JILA experiment}

The JILA experiment \cite{cornish_new} measured the observed number
of solitons $q$ and the total number of atoms $N$ in each case, and
the experimental data is presented in figure~\ref{fig8}(b) (points
with error bars). Note that both $q$ and $N$ were averaged over many
observations, and generally feature an error representing the spread
in their observations. Although this data shows the same trend as
our numerical data in figure~\ref{fig8}(a) (for fixed $a_{\rm s}$),
it should be note here that each point corresponds to a different
scattering length. We have therefore plotted the corresponding
numerical results in figure~\ref{fig8}(b). However, it is important
to note the following: the number of atoms measured experimentally
in the system $N_{\rm exp}$ will typically include non-condensed
atoms, whereas in our analysis $N$ describes the total number of
atoms in the BEC. In other words, these experimental and theoretical
parameters are not necessarily equal.

Initially we assume that all the experimentally observed atoms are
in the BEC, i.e. $N=N_{\rm exp}$ (black circles in
figure~\ref{fig8}(b)).  The numerical data is similar to the
experimental data with most numerical points lying within the spread
of the experimental observations. However for two data points the
numerical data over-estimates $q$, e.g. where the experiment
observes close to one and two solitons we predict two and three
solitons, respectively. However, assuming that $10\%$ of the atoms
observed experimentally represent non-condensed atoms, i.e. $N=0.9
N_{\rm exp}$ (red crosses in figure~\ref{fig8}(b)), we find that
these two anomalous points are corrected and every numerical point
is consistent with the experimental observations. This is a
conservative estimate for the number of non-condensed atoms given
the `hot' state of the condensate in the experiment.

Note that if there was zero phase difference between the solitons
they would overlap upon collision, and the critical number would
then be $N_{\rm c1}$ for $q\geq1$.

\section{Conclusions}
In this paper we have considered the solutions of an
attractively-interacting Bose-Einstein condensate in the mean-field
limit using the full Gross-Pitaevskii equation and variational
techniques.  Radial confinement is assumed throughout, and in the
axial direction we consider the absence of trapping and the presence
of confining or expulsive harmonic potentials. In particular we
evaluate the critical interaction parameter $k_{\rm c}$ for
condensate collapse across the whole range of axial geometries using
the full Gross-Pitaevskii equation.

We note that the variational method is not particularly sensitive to
whether the sech-based or gaussian-based ansatz is used, even in the
limits of zero and strong axial trapping. We find that the
variational method gives good qualitative agreement with the full
Gross-Pitaevskii equation, but that significant quantitative
differences can exist, particularly in the important critical points
for collapse and expansion.

In the absence of axial trapping, bright solitary wave solutions
exist, which are self-trapped in the axial direction by the
attractive interactions. According to the full solution of the
Gross-Pitaevskii equation, we find that bright solitary wave
solutions are stable to collapse only when the interaction parameter
$k<0.675\pm0.005$. Analysis of the energy `landscapes' reveals a
finite excitation energy to excite the bright solitary wave to
non-solitonic states, characterised by either collapse or dispersion
of the wavepacket. Unless the interaction parameter is close to the
collapse condition, i.e. $k>0.66$, the BSW will tend to decay by
dispersion. The excitation energy of a bright solitary wave is
proportional to the radial trap frequency, and for sufficiently weak
radial trapping, the excitation energy can be less than the typical
thermal energy, indicating thermodynamic instability of such states.

In the presence of either a confining or expulsive axial potential,
$k_{\rm c}$ becomes dependent on the trap geometry.  For the
confining potential, we regain the results of Gammal {\it et al.}
\cite{gammal} and good agreement with the experimental measurement
of $k_{\rm c}$ \cite{roberts,claussen}.  Under an expulsive axial
trap, there is also a {\em lower} critical interaction strength
$k_{\rm e}$ below which the trap leads to expansion of the
wavepacket. Therefore $k_{\rm c}$ and $k_{\rm e}$ define the upper
and lower boundaries of a range of solutions.  However, when the
trap ratio $|\lambda|$ exceeds a critical value $|\lambda|_{\rm
c}\approx 0.15$ (based on the full Gross-Pitaevskii equation), no
solutions exist for any interaction strength. Whereas for expulsive
potentials and no confinement the main decay channel is dispersion,
with axial confinement the only decay mechanism is collapse. Note
that, for a weak expulsive potential, one can tune the system such
that the energy barrier for collapse and dispersion are equal.

Finally, based on the JILA experiment \cite{cornish_new} we have
generated dynamical states of multiple `solitons' with alternating
phase. We have mapped out the critical numbers for these states and
find good agreement with the quantitative experimental data,
confirming the important role of $\pi$-phase differences in systems
of multiple solitons.

\ack We thank Y. S. Kivshar and T. J. Alexander for stimulating
discussions. We acknowledge financial support from the ARC
(NGP/AMM), University of Melbourne (NGP/AMM), UK EPSRC (CSA) and
Royal Society (SLC).


\begin{thebibliography}{99}
\bibitem{k_def} Under full 3D confinement we can define $k$ in terms of the
geometric mean of the harmonic oscillator lengths $a_{\rm 3D}=(a_z
a_r^2)^{1/3}=a_r \lambda^{-1/6}$, as used elsewhere
\cite{roberts,k_sph_sym,gammal,gammal2,yukalov}.
\end{thebibliography}
\end{document}